\documentclass[12pt]{article}

\ifx\pdfoutput\undefined
\usepackage[dvips,bookmarks]{hyperref}
\else
\usepackage{hyperref}
\fi
\hypersetup{colorlinks=false,bookmarksopen,bookmarksnumbered,citecolor=blue,
   pdfstartview=FitH}

\usepackage{latexsym}
\usepackage{amssymb,amsfonts,amsmath}
\usepackage{graphicx} 
\usepackage{indentfirst}
\usepackage{bbm}
\usepackage{amssymb}
\usepackage{verbatim}
\usepackage{amsmath, amsthm,amssymb}
\usepackage{mathrsfs}
\usepackage{hyperref}
\usepackage{amsfonts}
\usepackage{dsfont}
\usepackage{slashed, tensor}
\usepackage{booktabs}
\usepackage{graphicx}
\usepackage{mathrsfs}
\parskip=\medskipamount

\usepackage[mathscr]{euscript} % alternate script style

\newcommand {\cC}{{\cal C}}

\newcommand {\cF}{{\cal F}}

\newcommand {\cN}{{\cal N}}

\newcommand {\cW}{{\cal W}}

\def\a{\alpha}
\def\b{\beta}

\def\d{\delta}

\def\g{\gamma}
\def\G{\Gamma}

\def\l{\lambda}

\def\q{\theta}

\def\z{\zeta}
\def\D{\Delta}
\def\F{\Phi}
\def\J{\Psi}

\def\ri{{\rm i}}

\newcommand{\pa}{\partial}
\newcommand{\hf}{\frac12}

\newcommand{\vf}{\varphi}

\newcommand{\be}{\begin{equation}}
\newcommand{\ee}{\end{equation}}
\newcommand{\bea}{\begin{eqnarray}}
\newcommand{\eea}{\end{eqnarray}}
\newcommand{\non}{\nonumber}
\newcommand{\ba}{\begin{array}}
\newcommand{\ea}{\end{array}}

\def\double #1{#1{\hbox{\kern-2pt $#1$}}}

\newcommand{\bsubeq}{\begin{subequations}}
\newcommand{\esubeq}{\end{subequations}}

\newcommand{\rd}{\mathrm d}
\numberwithin{equation}{section}  % Resets equation number at beginning of each section

\newcommand{\pde}{\partial}

\begin{document}
%%%%%%%%%%%%%%%%
%%%%%%%%%%%%%%%

\begin{titlepage}
\begin{flushright}
June, 2016 \\
Revised version: September, 2016
\end{flushright}
\vspace{5mm}

\begin{center}
{\Large \bf 
Higher spin super-Cotton tensors and generalisations of the linear-chiral duality 
 in three dimensions}
\end{center}

\begin{center}

{\bf
Sergei M. Kuzenko
} \\
\vspace{5mm}

\footnotesize{
{\it School of Physics M013, The University of Western Australia\\
35 Stirling Highway, Crawley W.A. 6009, Australia}}  
~\\

\vspace{2mm}
~\\
\texttt{sergei.kuzenko@uwa.edu.au 
}\\
\vspace{2mm}

\end{center}

\begin{abstract}
\baselineskip=14pt

In three spacetime dimensions, (super)conformal geometry is controlled by the
(super-)Cotton tensor. We present a new duality transformation 
for $\cN$-extended supersymmetric theories  formulated in terms of 
the linearised super-Cotton tensor or its 
higher spin extensions for the cases $\cN=2,\,1,\,0$. In the $\cN=2$ case, 
this transformation is a generalisation of the linear-chiral duality, 
which  
%is known to 
provides a dual description in terms of
chiral superfields for general models of self-interacting  $\cN=2$ vector 
multiplets in three dimensions and $\cN=1$ tensor multiplets in four dimensions. 
For superspin-1 (gravitino multiplet), superspin-3/2 (supergravity multiplet) 
and any higher superspin $s\geq 2$, the duality transformation relates a higher-derivative theory 
to one containing at most two derivatives at the component level.
In the $\cN=1$ case, we introduce gauge prepotentials
for higher spin superconformal gravity and construct the corresponding 
super-Cotton tensors, as well as the higher spin extensions of the linearised $\cN=1$ conformal  
supergravity action. Our $\cN=1$ duality transformation is a higher spin extension 
of the known superfield duality relating the massless $\cN=1$ vector and scalar multiplets. 
%In the non-supersymmetric ($\cN=0$) case,  the gauge prepotentials
%for higher spin conformal geometry (both bosonic and fermionic) and the corresponding 
%Cotton tensors can be obtained from their $\cN=1$ counterparts
%by carrying out $\cN=1 \to \cN=0$ reduction. 
%In the bosonic sector, this reduction is shown 
%to lead to the Cotton tensors for higher spins constructed by Pope and Townsend, 
%and by Damour and Deser in the spin-3 case.
Our $\cN=0$ duality transformation is a higher spin extension 
of the vector-scalar duality.
\end{abstract}

\vfill

\vfill
\end{titlepage}

\newpage
\renewcommand{\thefootnote}{\arabic{footnote}}
\setcounter{footnote}{0}

%\tableofcontents

%%%%%%%%%%%%%%%%%%%%%%%%%%%%%%%%%%%%%%%%%%%%%%%%%%%%%%
%%%%%%%%%%%%%%%%%%%%%%%%%%%%%%%%%%%%%%%%%%%%%%%%%%%%%%

\allowdisplaybreaks

\section{Introduction}

Recently there has been renewed interest in dualities in three-dimensional (3D)
field theories \cite{KT,MN,SSWW}.
In this paper we consider  3D duality transformations for theories involving higher spin analogues 
of the Cotton tensor or its supersymmetric generalisations -- the $\cN=1$ and $\cN=2$ 
super-Cotton tensors \cite{K12,KT-M12,BKNT-M}. The specific feature of 3D conformal gravity is that its geometry can be formulated such that the Cotton tensor fully determines the algebra
of covariant derivatives, see e.g. \cite{BKNT-M}. Similarly, in 3D $\cN$-extended conformal supergravity formulated in conformal superspace \cite{BKNT-M}, 
the corresponding superspace geometry is controlled by the super-Cotton tensor. 
In the $\cN=2$ case, our higher spin duality transformation 
may be thought of as a generalisation of the famous linear-chiral duality.

The linear-chiral duality \cite{Siegel,LR} is of fundamental 
importance in supersymmetric field theory, supergravity and string theory, 
in particular in the context of supersymmetric nonlinear sigma models 
\cite{LR,HKLR,LR-projective1}. It provides a dual description in terms of
chiral superfields for general models of self-interacting 3D $\cN=2$ vector 
multiplets or 4D $\cN=1$ tensor multiplets \cite{Siegel}. The only assumption 
for the duality to work is that the 3D $\cN=2$ vector 
multiplet or 4D $\cN=1$ tensor multiplet appears in 
the superfield Lagrangian, $L(W)$, only via its field strength $W$, 
which is a real linear superfield, 
\bea
\bar D^2 W = 0~, \qquad W=\bar W \quad \Longrightarrow \quad
D^2 W =0~.
\label{1.1}
\eea 

It is pertinent to recall the definition of the linear-chiral duality in the 
3D $\cN=2$ case we are interested in. We start from a  self-interacting 
vector multiplet model with action 
\bea
S[W] = 
\int  {\rm d}^3x {\rm d}^2 \q \rd^2 \bar \q \, L(W) ~, 
\label{1.2}
\eea
and associate with it the following first-order model 
\bea
S[\cW, \J, \bar \J] = \int 
{\rm d}^3x {\rm d}^2 \q \rd^2 \bar \q \, 
\Big\{ L(\cW) -  (\J + \bar \J )\cW\Big\} ~, 
\qquad \bar D_\a \J=0~.
\label{1.3}
\eea
Here the dynamical variables are a real unconstrained superfield $\cW$, 
a chiral scalar $\J$ and its complex conjugate $\bar \J$. 
Varying $S[\cW, \J, \bar \J]$ with respect to the Lagrange multiplier $\J$ gives 
the equation of motion $\bar D^2 \cW = 0$, and hence $\cW=W$. 
Then the second term on the right of $S[\cW, \J, \bar \J]$ drops out, and we are back to the vector multiplet model \eqref{1.2}. 
On the other hand, we can vary \eqref{1.3} with respect to 
$\cW$ resulting in the equation of motion 
\bea
L'(\cW) = \J + \bar \J ~.
\eea
This equation allows us to express $\cW$ as a function of $\J$ and $\bar \J$, 
and then \eqref{1.3} turns into the dual action 
\bea
S_{\rm D} [ \J, \bar \J] = \int 
{\rm d}^3x {\rm d}^2 \q \rd^2 \bar \q \, 
L_{\rm D}(\J, \bar \J) ~.
\eea
It remains to point out that 
the constraint \eqref{1.1} is solved in the 3D case by 
\bea
W= \D H~, \qquad  \D =\frac{\ri}{2} D^\a \bar D_\a ~,
\label{166}
\eea
where the real prepotential $H$ is defined modulo gauge transformations
of the form
\bea
\d H = \l +\bar \l ~, \qquad \bar D_\a \l=0~.
\label{177}
\eea

It is worth recalling one 
more type of duality that is naturally defined
in the 3D $\cN=1$ and  4D $\cN=2$ cases, 
the so-called complex linear-chiral duality \cite{GS1,GS2}
(see also \cite{GGRS} for a review). 
It provides a dual description in terms of
chiral superfields for general models of self-interacting complex 
linear superfields $\G$ and their conjugates $\bar \G$. 
The complex 
linear-chiral duality plays a fundamental role in the context 
of off-shell supersymmetric sigma models with eight supercharges 
\cite{LR-projective1,GK}.
The complex linear superfield $\G$ is defined by the only constraint
\bea
\bar D^2 \G =0~.
\eea
The complex linear-chiral duality works as follows. Consider 
a 3D $\cN=2$ supersymmetric field with action 
\bea
S[\G, \bar \G] =
\int 
{\rm d}^3x {\rm d}^2 \q \rd^2 \bar \q \,  L(\G, \bar \G)~.
\eea
We associate with it a first-oder action of the form
\bea
S[V, \bar V , \J, \bar \J]=  
\int 
{\rm d}^3x {\rm d}^2 \q \rd^2 \bar \q \, 
\Big\{ L(V, \bar V)
- \J V - \bar \J \bar V  \Big\}~, \qquad \bar D_\a \J =0~.
\label{first-order-action}
\eea
Here the dynamical superfields comprise a complex unconstrained 
scalar $V$, a chiral scalar $\J$ and their conjugates. 
Varying \eqref{first-order-action}
with respect to the Lagrange multiplier $\J$ gives $V = \G$, and then the second
term in \eqref{first-order-action} drops out as a consequence of the 
identities
\bea
\int {\rm d}^3x {\rm d}^2 \q \rd^2 \bar \q \, U 
= -\frac{1}{4} \int {\rm d}^3x {\rm d}^2 \q \, \bar D^2 U
= -\frac{1}{4} \int {\rm d}^3x {\rm d}^2 \bar \q \,  D^2 U
~,
\eea
for any superfield $U$.
As a result, the first-order action 
reduces to the original one, $S[\G, \bar \G]$.
On the other hand, we can 
consider the equation of motion for $V$, 
\bea
\frac{\pa}{\pa V} L(V, \bar V) = \J ~,
\eea
and its conjugate. The latter equations allow us to express the auxiliary 
superfields $V$ and $\bar V$ in terms of $\J$ and $\bar \J$. 
Then  \eqref{first-order-action} turns into the dual action 
\bea
S_{\rm D}[\J, \bar \J] =
\int {\rm d}^3x {\rm d}^2 \q \rd^2 \bar \q \,  
L_{\rm D}(\J, \bar \J)~.
\eea

It should be mentioned that the complex linear-chiral duality 
can also be introduced in the reverse order, 
by starting with a chiral model 
\bea
S[\J, \bar \J] =
\int  {\rm d}^3x {\rm d}^2 \q \rd^2 \bar \q \,  
K(\J, \bar \J)~, \qquad \bar D_\a \J =0~,
\eea
and then applying a Legendre transformation to $S[\J, \bar \J]$ in order 
to result in a model described by a complex linear superfield $\G$ and 
its conjugate $\bar \G$. This makes use of the first-order action
\bea
S[U, \bar U , \G, \bar \G]=  
\int 
{\rm d}^3x {\rm d}^2 \q \rd^2 \bar \q \, 
\Big\{ K(U, \bar U)
- \G U - \bar \G \bar U  \Big\}~, \qquad \bar D^2 \G =0~.
\label{1.15}
\eea

In the 4D $\cN=1$ case, the first-order action \eqref{1.15} with 
$K(U, \bar U) =U\bar U$
was considered for the first time by Zumino \cite{Zumino1980}.
However, he did not realise the fact that this construction leads
to a new off-shell description for the scalar multiplet, 
which was an observation made in  \cite{GS1,GS2}.

The higher-spin generalisation of the complex linear-chiral duality 
has been given in \cite{KPS,KS93,KS94,KO}. For every integer
$2s=3,4, \dots$, it relates the two off-shell formulations for the massless superspin-$s$
multiplet in four dimensions constructed\footnote{See \cite{BK}
for a review of the models proposed in   \cite{KPS,KS93}.}
in \cite{KPS,KS93,KS94}
and for the massive superspin-$s$ multiplet in three dimensions 
presented in  \cite{KO}.
The goal of this paper is to give  higher-spin 
generalisations of the 3D $\cN=2$ linear-chiral duality in three dimensions
and its $\cN=1$ and $\cN=0$ cousins.

This paper is organised as follows. The $\cN=2$ duality transformation is 
presented in section 2. In section 3  we introduce gauge prepotentials
for higher spin superconformal geometry,  construct the corresponding 
super-Cotton tensors, and present the higher spin extension of the linearised $\cN=1$ conformal  supergravity action. Our $\cN=1$ duality transformation is also 
described in this section. Finally, section 4 is devoted to the non-supersymmetric
($\cN=0$) higher spin duality transformation.

%%%%%%%%%%%%%%%%%%%%%%%%%%%
%%%%%%%%%%%%%%%%%%%%%%%%%%%

\section{$\cN=2$ duality}

Let $n$ be  a positive integer.  
We recall the higher-spin $\cN=2$ superconformal  field strength, 
$W_{\a(n)} = \bar W_{\a(n)}$, introduced in \cite{KO}
\bea
&&W_{\a_1 \dots \a_n}  (H)
:= \frac{1}{2^{n-1}} 
\sum\limits_{J=0}^{\left \lfloor{n/2}\right \rfloor}
\bigg\{
\binom{n}{2J} 
\Delta  \Box^{J}\pde_{(\a_{1}}{}^{\b_{1}}
\dots
\pde_{\a_{n-2J}}{}^{\b_{n-2J}}H_{\a_{n-2J+1}\dots\a_{n})\b_1 \dots\b_{n-2J}}~~~~
\nonumber \\
&&\qquad \qquad +
\binom{n}{2J+1}\Delta^{2}\Box^{J}\pde_{(\a_{1}}{}^{\b_{1}}
\dots\pde_{\a_{n-2J -1}}{}^{\b_{n-2J -1}}H_{\a_{n-2J}\dots\a_{n})
\b_1 \dots \b_{n-2J -1} }\bigg\}~,~~~~~
\label{eq:HSFSUniversal}
\eea
where $\left \lfloor{x}\right \rfloor$ denotes the floor (or the integer part) of 
a number $x$. The field strength $W_{\a(n)}$ is a descendant of 
 the real unconstrained prepotential $H_{\a(n)}$ defined 
 modulo gauge transformations of the form
\begin{subequations}\label{5.19}
\bea
\d H_{\a(n)} = g_{\a(n)} + \bar g_{\a(n)}~, \qquad
g_{\a_1\dots \a_n} = \bar{D}_{(\a_{1}}L_{\a_{2}...\a_{n})} ~,
\eea
where the complex gauge parameter  $g_{\a(n)}  $ 
is an arbitrary longitudinal linear superfield, 
\bea
\bar{D}_{(\a_{1}}g_{\a_{2}...\a_{n+1})}  =0~.
\eea
\end{subequations}
 The field strength is invariant under the gauge transformations \eqref{5.19}, 
 \bea
  \d W_{\a(n)}=0~, 
\eea
and  obeys the Bianchi identity
 \bea
\bar D^{\b}W_{\b\a_1 \dots \a_{n-1}}=0 \quad \Longleftrightarrow \quad
{D}^{\b}W_{\b\a_1 \dots \a_{n-1}}=0 ~,
\label{2.4}
\eea
which implies 
\bea
\bar D^2 W_{\a(n)} =  D^2 W_{\a(n)} =0~.
\label{266}
\eea 
As demonstrated in \cite{KO},  $W_{\a(n)}$ is a primary superfield of dimension 
 $(1+n/2)$ if the prepotential $H_{\a(n)}$ is chosen to be 
 primary of  dimension $(-n/2)$.
 Associated with $W_{\a(n)}$ is the $\cN=2$ superconformal Chern-Simons 
 action \cite{KO}
\begin{equation}
S_{\rm CS} [H]=  \ri^n \int 
{\rm d}^3x {\rm d}^2 \q \rd^2 \bar \q 
\, H^{\a(n)}W_{\a(n)} (H)~,
\label{eq:HSChernSimmonsAction}
\end{equation}
which is invariant under the gauge transformations \eqref{5.19}.
In the $n=0$ case, the Bianchi identity \eqref{2.4} should be replaced with \eqref{1.1}. 

In the $n=2$ case, the field strength $W_{\a\b}(H)$ coincides with the linearised
version \cite{CDFKS,KO} of the $\cN=2$ super-Cotton tensor 
\cite{K12,BKNT-M}. Thus the field strength \eqref{eq:HSFSUniversal} for $n>2$
is the higher-spin extension of 
the super-Cotton tensor.

For $n=0$  the Bianchi identity \eqref{2.4} is not defined, 
and is instead replaced with its corollary 
\eqref{266}. In this case, the expression \eqref{eq:HSFSUniversal} reduces to the 
vector multiplet field strength \eqref{166}, and the gauge invariance \eqref{5.19}
turns into \eqref{177}. Finally, for $n=0$  the action \eqref{eq:HSChernSimmonsAction}
reduces to the topological mass term for the Abelian vector multiplet 
\cite{Siegel2,ZP,Ivanov91}. 

Of crucial importance for our analysis is the fact  that 
the expression \eqref{eq:HSFSUniversal}
 is the general solution to the constraint \eqref{2.4}.
 This observation is the key to introducing a new type of duality.
Let us consider a higher-derivative  $\cN=2$ superconformal  theory with  action
\bea
S[W ]= \int {\rm d}^3x {\rm d}^2 \q \rd^2 \bar \q 
\, \F \bar \F \,
L \left( \frac{W_{\a(n)} }{(\F \bar \F)^{1+ n/2}} \right)~,
\label{2.7}
\eea 
where the superconformal compensator $\F$ 
is a  nowhere vanishing primary superfield of dimension 1/2.
The origin of $\F$ is not important for us. In particular, 
$\F$ may be frozen to a constant value, 
and then we result with a theory described solely in terms of the higher-spin 
gauge superfield $H_{\a(n)}$.
We can associate with \eqref{2.7} the following 
first-order model 
\bea
S[\cW , G, \bar G]=  \int {\rm d}^3x {\rm d}^2 \q \rd^2 \bar \q 
\left\{
\F \bar \F \,
L \left( \frac{\cW_{\a(n)} }{(\F \bar \F)^{1+ n/2}} \right)
- \ri^n( G^{\a(n)} + \bar G^{\a(n) } ) \cW_{\a (n)} \right\} ~,~~~~~
\label{2.8}
\eea 
where $\cW_{\a(n)} $ is a real unconstrained  superfield, while 
$G_{\a(n)}$ is a longitudinal linear superfield, 
\bea
\bar D_{(\a_1} G_{\a_2 \dots \a_{n+1})} =0~.
\label{LL}
\eea
The general solution of this constraint is 
\bea
G_{\a_1 \dots \a_n} &=& \bar D_{(\a_1} \z_{\a_2 \dots \a_n)} \ ,
\eea
for some unconstrained complex  prepotential $\z_{\a(n-1)}$.
Varying \eqref{2.8} with respect to $G^{\a(n)}$ and its conjugate gives
 \bea
\bar D^{\b}\cW_{\b\a_1 \dots \a_{n-1}}=0 \quad \Longleftrightarrow \quad
{D}^{\b}\cW_{\b\a_1 \dots \a_{n-1}}=0 ~,
\eea
which means that $\cW_{\a(n)} = W_{\a(n)}$. 
Plugging this in $S[\cW , G, \bar G]$,
the second term in \eqref{2.8}
drops out, and we return to the original action \eqref{2.7}.
On the other hand, we can start from the first-order model \eqref{2.8} 
and integrate out the auxiliary superfield $\cW_{\a(n)}$.
This leads to a dual action of the form
\bea
S_{\rm D}[G, \bar G]=  \int {\rm d}^3x {\rm d}^2 \q \rd^2 \bar \q \,
\F \bar \F \,
L _{\rm D}\Big( ( G^{\a(n)} + \bar G^{\a(n) } ) (\F \bar \F)^{n/2}\Big)~.
\label{2.12}
\eea
Unlike the original action \eqref{2.7}, its dual $S_{\rm D}[G, \bar G]$ does not contain higher derivatives.

In describing our duality transformation, we assumed $n>0$. It is easy to see that 
it reduces to the linear-chiral duality in the $n=0$ case.

%%%%%%%%%%%%%%%%
%%%%%%%%%%%%%%%%%

\section{$\cN=1$ duality} 

The results in the previous section can be used to obtain a new type
of $\cN=1$ duality  by carrying out the $\cN=2 \to \cN=1$ superspace 
reduction sketched in \cite{KO}.
Applying this reduction to 
\eqref{eq:HSFSUniversal} 
leads to the higher-spin $\cN=1$ superconformal  field strength\footnote{See 
\cite{KTsulaia} for a detailed derivation.}
\bea
&&W_{\a_1 \dots \a_n}  (H)
:= \frac{1}{2^{n-1}} 
\sum\limits_{J=0}^{\left \lfloor{n/2}\right \rfloor}
\bigg\{
\binom{n}{2J}  \Box^{J}\pde_{(\a_{1}}{}^{\b_{1}}
\dots
\pde_{\a_{n-2J}}{}^{\b_{n-2J}}H_{\a_{n-2J+1}\dots\a_{n})\b_1 \dots\b_{n-2J}}~~~~
\nonumber \\
&&\qquad \qquad -\frac{\ri}{2} 
\binom{n}{2J+1}D^{2}\Box^{J}\pde_{(\a_{1}}{}^{\b_{1}}
\dots\pde_{\a_{n-2J -1}}{}^{\b_{n-2J -1}}H_{\a_{n-2J}\dots\a_{n})
\b_1 \dots \b_{n-2J -1} }\bigg\}~,~~~~~
\label{3.1}
\eea
which is real, $W_{\a(n)} = \bar W_{\a(n)}$. 
The field strength $W_{\a(n)}$ is a descendant of 
 the real unconstrained prepotential $H_{\a(n)}$ defined 
 modulo gauge transformations of the form 
 \bea
\d H_{\a(n)} = \ri^n {D}_{(\a_{1}}\z_{\a_{2}...\a_{n})} ~,\qquad 
\bar \z_{\a(n-1)}=\z_{\a(n-1)}~.
\label{3.2}
\eea
 The field strength is invariant under the gauge transformations \eqref{3.2}, 
 \bea
  \d W_{\a(n)}=0~, 
\eea
and  obeys the Bianchi identity
\bea
D^{\b}W_{\b\a_1 \dots \a_{n-1}}=0  ~.
\label{3.3}
\eea
It may be shown that  $W_{\a(n)}$ is a primary superfield of dimension 
 $(1+n/2)$, in the sense of \cite{KPT-MvU,BKS},
 if the prepotential $H_{\a(n)}$ is  primary of 
 dimension $(1-n/2)$.
 Associated with $W_{\a(n)}$ is the $\cN=1$ superconformal Chern-Simons 
 action
\begin{equation}
S_{\rm CS} [H]=  \ri^{n-1} \int 
{\rm d}^3x {\rm d}^2 \q 
\, H^{\a(n)}W_{\a(n)} (H)~,
\label{355}
\end{equation}
which is invariant under the gauge transformations \eqref{3.2}.
The superconformal invariance of $S_{\rm CS} [H]$ is discussed in detail in 
\cite{KTsulaia}. The action \eqref{355} coincides for $n=1$ with the topological mass term for the Abelian vector multiplet  \cite{Siegel2}. In the $n=3$ case, \eqref{355}
proves to be the linearised action for $\cN=1$ conformal supergravity
\cite{GGRS,KT-M12}.

For $n=1$ the field strength \eqref{3.1} is 
\bea
W_\a  = -\pa_\a{}^\b H_\b +\frac{\ri}{2} D^2 H_\a = \ri D^\b D_\a H_\b~,
\label{3.4}
\eea
as a consequence of the anti-commutation relation 
\bea
\{D_\a, D_\b \} = 2\ri \pa_{\a\b} ~.
\eea
The final expression for $W_\a$ in \eqref{3.4}
coincides with the gauge-invariant field strength 
of a vector multiplet \cite{GGRS}. The Bianchi identity 
$D^\a W_\a=0$ is a corollary of 
\bea
D^\a D_\b D_\a =0 \quad \Longrightarrow \quad 
[D_\a D_\b, D_\g D_\d ]=0~. 
\label{3.5}
\eea
For $n=2$ the field strength \eqref{3.1} can be seen to coincide with the gravitino field 
strength \cite{GGRS}. Finally, for  $n=3$ the field strength \eqref{3.1} is
the linearised version  \cite{KNT-M} of the $\cN=1$ super-Cotton tensor 
\cite{KT-M12,BKNT-M}. This is why \eqref{3.1} can be called the higher spin super-Cotton tensor.

It should be pointed out that \eqref{3.1} is the general solution of the constraint
\eqref{3.3}. The simplest way to prove this is the observation that 
the field strength \eqref{3.1} may be recast in the form
\bea
W_{\a(n)} \propto \ri^n D^{\b_1} D_{\a_1} \dots D^{\b_n} D_{\a_n} H_{\b_1 \dots \b_n}~.
\label{3.9}
\eea
It is completely symmetric, $W_{\a_1 \dots \a_n} =W_{(\a_1 \dots \a_n)}$,
 as a consequence of \eqref{3.5}.

Let us consider a higher-derivative  $\cN=1$ superconformal  theory with  action
\bea
S[W] = \ri \int {\rm d}^3x {\rm d}^2 \q \, \vf^4 L \left(
\frac{W_{\a(n)} }{ \vf^{n+2} } \right)~,
\label{3.10}
\eea
where $\vf$ is a real conformal compensator of dimension 1/2.
This model possesses a dual description. Indeed, 
we can associate with \eqref{3.10} the following first-order model 
\bea
S[\cW, G] = \ri \int {\rm d}^3x {\rm d}^2 \q \, \left\{
\vf^4 L \left(
\frac{\cW_{\a(n)} }{ \vf^{n+2} } \right)
-\ri^n G^{\a(n)} \cW_{\a(n)}
\right\}
~,
\label{3.11}
\eea
where $\cW_{\a(n)}$ is an unconstrained superfield, and the Lagrange multiplier 
has the form
\bea
G_{\a(n)} = \ri^n {D}_{(\a_{1}}\J_{\a_{2}...\a_{n})} ~,\qquad 
\bar \J_{\a(n-1)}=\J_{\a(n-1)}~.
\eea
Varying $S[\cW, G] $ with respect to the Lagrange $\J_{\a(n-1)}$ leads us back to the original action \eqref{3.10}, and therefore the models \eqref{3.10} and \eqref{3.11}
are equivalent. On the other hand, we can integrate out the auxiliary superfield 
$\cW_{\a(n)}$ from \eqref{3.11}, which leads us to 
a dual action of the form
\bea
S[ G] = \ri \int {\rm d}^3x {\rm d}^2 \q \, 
\vf^4 L_{\rm D} \left(\vf^{n-2} G_{\a(n)}    \right)~.
\eea
Unlike the original action \eqref{3.10}, which is a higher-derivative theory for $n>1$, 
its dual $S_{\rm D}[G]$ is free of higher derivatives.
In the $n=1$ case, the duality transformation described corresponds to the standard 
duality between the $\cN=1$ scalar and vector multiplets in three dimensions
\cite{HKLR}.

%%%%%%%%%%%%%%%%%%%%%%%
%%%%%%%%%%%%%%%%%%%%%%%%

\section{$\cN=0$ duality}

The $\cN=1$ higher spin super-Cotton tensor and the 
$\cN=1$ duality transformation described in section 3 were obtained by performing 
the $\cN=2 \to \cN=1$ superspace reduction, in analogy 
with the earlier results for extended supersymmetric nonlinear sigma models 
\cite{KLT-M,BKT-M}.
 Actually one can continue this process one step further 
and carry out the $\cN=1 \to \cN=0$ reduction.
This gives 
%(up to a factor) 
the higher-spin conformal field strength
\bea
&&C_{\a(n)}  (h)
:=\frac{1}{2^{n-1}} \sum\limits_{J=0}^{\left \lfloor{n/2}\right \rfloor}
\binom{n}{2J+1}
\Box^{J}\pde_{(\a_{1}}{}^{\b_{1}}
\dots\pde_{\a_{n-2J -1}}{}^{\b_{n-2J -1}}h_{\a_{n-2J}\dots\a_{n})
\b_1 \dots \b_{n-2J -1} }~,~~~~~
\label{4.1}
\eea
which is defined for $n \geq 2$.
It is a descendant of the real prepotential $h_{\a(n)}(x)$ defined modulo gauge 
transformations of the form
\bea
\d h_{\a(n)} = \pa_{(\a_1\a_2} \z_{\a_3 \dots \a_n)}~.
\label{4.2}
\eea
The field strength \eqref{4.1} is invariant under these gauge transformations, 
\bea
\d C_{\a(n)} =0~,
\eea
and obeys the Bianchi identity
\bea
\pa^{\b\g} C_{\b\g \a_1 \dots \a_{n-2}} =0~.
\label{4.4}
\eea
It may be shown that  $C_{\a(n)}$ is a primary field of dimension 
 $(1+n/2)$ if the prepotential $h_{\a(n)}$ is  primary of 
 dimension $(2-n/2)$. Associated with $C_{\a(n)}$ is the conformal
 Chern-Simons action \cite{PopeTownsend}
 \bea
 S_{\rm CS} [h] =\ri^n  \int \rd^3 x\, h^{\a(n)} C_{\a(n)} (h)~,
\eea
which is invariant under the gauge transformations \eqref{4.2}.
 
 In the case of even rank, $n=2s$, with $s=1,2,\dots$, the field strength \eqref{4.1} 
 coincides with the bosonic higher spin Cotton tensor given originally by Pope and Townsend 
 \cite{PopeTownsend}. It reduces to the linearised Cotton tensor for $n=4$, 
 and to the Maxwell field strength for  $n=2$.
It should be pointed out that the conformal spin-3 case, $n=6$, 
was studied for the first time in \cite{DD}.
In the case of odd rank, $n=2s+1$,
eq.  \eqref{4.1} describes fermionic higher spin conformal field strengths.
They did not appear in \cite{PopeTownsend}. The spin-3/2 case, $n=3$,  
was considered in \cite{ABdeRST}, where the field strength $C_{\a(3)}$ was called the
Cottino tensor.

The field strength \eqref{4.1} proves to be the general solution to the conservation equation
\eqref{4.4}. This result has recently been proved in \cite{HHL} in the bosonic case, 
$n=2s$,
and the proof given is quite nontrivial. There is an alternative proof based on supersymmetry considerations. The point is that the higher spin Cotton tensor $C_{\a(n)}$ may be imbedded
into the $\cN=1$ super-Cotton tensor $W_{\a(n)}$ as its lowest ($\q$-independent) component.
The latter obeys the constraint \eqref{3.3}, which has the general solution 
\eqref{3.1} or, equivalently, \eqref{3.9}. The fact that \eqref{3.9} is the general solution to 
\eqref{3.3}, is a corollary of the $\cN=1$ identities \eqref{3.5}.

An important feature of the spin-$\frac{n}{2}$ Cotton tensor \eqref{4.1} 
for $n>1$ is that it can 
be represented as a linear superposition of the 
%gauge-invariant 
equations of motion 
for the massless spin-$\frac{n}{2}$ 
Fronsdal 
action in three dimensions \cite{Fronsdal,FF} (see \cite{KO} for a review),
with the coefficients being linear higher-derivative operators.
A similar result in the $\cN=2$ supersymmetric case was spelt out 
in detail in \cite{KO}, which is why
here our consideration will be restricted to the bosonic case,  $n=2s$, 
with $s= 2,3, \dots$
The massless spin-$s$ action is described by two real gauge fields
$\vf^i =\big\{h_{\a(2s)}\, ,  \,h_{\a(2s-4)}\big\}$
defined modulo the gauge transformations 
%\begin{subequations} \label{B.2}
\bea
\delta h_{\a(2s)}&=&\pa_{(\a_{1}\a_{2}}\zeta_{\a_{3} \dots \a_{2s}) } ~,\qquad 
\delta h_{\a(2s-4)}=
\frac{1}{2s-1}
\pa^{\b \g }\zeta_{\b \g \a_1 \dots \a_{2s-4} }~.
%\label{B.2b}
\eea
The field $h_{\a(2s)}$ is the conformal spin-$s$ prepotential, 
while the other field $h_{\a(2s-4)}$ is a gauge compensator.
Associated with the gauge fields $h_{\a(2s)} $ and $ h_{\a(2s-4)}$ 
%One may construct 
are
the following gauge-invariant field strengths
\begin{subequations} 
\bea
\cF_{\a(2s)}&=&\Box h_{\a(2s)}
+\hf s\pde^{\b(2)}\pde_{\a(2)}h_{\a(2s-2)\b(2)}
-\hf (s-1)(2s-3)\pde_{\a(2)}\pde_{\a(2)} h_{\a(2s-4)}~, ~~~~~~\label{B.4a}\\
\cF_{\a(2s-4)}&=&\pde^{\b(2)}\pde^{\gamma(2)}h_{\b(2)\gamma(2)\a(2s-4)}
+8 \frac{(s-1)}{s}\Box h_{\a(2s-4)} \non \\
&&\qquad \qquad \qquad 
+(s-2)(2s-5) \pde^{\b(2)}\pde_{\a(2)}h_{\a(2s-6)\b(2)} ~,\label{B.4b}
\eea
\end{subequations}
which are related to each other by the Bianchi identity 
\bea
\pa^{\b\g} \cF_{\a_1 \dots \a_{2s-2} \b\g} 
= \frac{(s-1) (2s-3)}{2(2s-1)} \pa_{(\a_1 \a_2} \cF_{\a_3 \dots \a_{2s-2})}~.
\eea
In terms of these field strengths, the equations of motion 
for  the massless spin-$s$ field are $\cF_{\a(2s)}=0$ and $\cF_{\a(2s-4)}=0$
(compare with eq. (B.4) in \cite{KO}).
We state that the spin-$s$ Cotton tensor $C_{\a(2s)}$ can be expressed in terms of
$\cF_{\a(2s)}$ and $\cF_{\a(2s-4)}$, in particular for the spin-2 and spin-3 cases we 
have
\begin{subequations}
\bea
C_{\a(4)} &=& 
%{8} 
\pa_{(\a_1}{}^\b \cF_{\a_2 \a_3 \a_4)\b} ~, \label{4.9a}\\
C_{\a(6)} &=& 
%32 
\pa_{(\a_1}{}^{\b_1} \pa_{\a_2}{}^{\b_2} \pa_{\a_3}{}^{\b_3}
\cF_{\a_1\a_2\a_3 )\b_1\b_2\b_2}
-\frac{9}{80} \pa_{(\a_1}{}^\b \pa_{\a_2\a_3} \pa_{\a_4\a_5} \cF_{\a_6) \b}~.
\label{4.9b}
\eea
\end{subequations}
The general result for $s>3$ may be deduced, e.g., 
from the relation (6.25) in \cite{KO}
by reducing it to components.
%performing the $\cN=2 \to \cN=0$ superspace reduction.
There is a simple explanation why  $C_{\a(2s)}$ can be expressed in terms of
$\cF_{\a(2s)}$ and $\cF_{\a(2s-4)}$. 
It is based on the well-known result that on 
%As is well known, 
the equations of motion  $\cF_{\a(2s)}=0$ and $\cF_{\a(2s-4)}=0$
%imply that 
the fields  $h_{\a(2s)} $ and $ h_{\a(2s-4)}$ can be completely 
gauged away  (see \cite{KO} for a review and proof), and hence $C_{\a(2s)}=0$.
This implies that $C_{\a(2s)}$ is a descendant of
$\cF_{\a(2s)}$ and $\cF_{\a(2s-4)}$. 
%Eq. \eqref{4.9b} immediately leads to the spin-3 results of \cite{BHT}.
Eq. \eqref{4.9a} is equivalent to the standard linearised expression for the Cotton 
tensor in terms of the Schouten tensor. In \cite{BHT} a spin-3 
analog of the Schouten tensor was introduced as a potential for $C_{\a(6)}$. 
This spin-3 result of \cite{BHT} readily  follows from \eqref{4.9b}.
In our opinion, the higher spin generalisation of the spin-2 relation \eqref{4.9a}
is the expression for $C_{\a(2s)}$ in terms of $\cF_{\a(2s)}$ and $\cF_{\a(2s-4)}$, 
see \cite{KO} for the $\cN=2$ supersymmetric case.

Let us turn to introducing the aforementioned $\cN=0$ duality transformation.
We consider a higher-derivative  conformal  theory with  action
\bea
S[C] =  \int {\rm d}^3x   \, \vf^6 L \left(
\frac{C_{\a(n)} }{ \vf^{n+2} } \right)~,
\label{4.6}
\eea
where $\vf$ is a real conformal compensator of dimension 1/2.
To obtain a dual description for the model, 
we can associate with \eqref{4.6} the following first-order model 
\bea
S[\cC, G] =  \int {\rm d}^3x \, \left\{
\vf^6 L \left(
\frac{\cC_{\a(n)} }{ \vf^{n+2} } \right)
-\ri^n G^{\a(n)} \cC_{\a(n)}
\right\}
~,
\label{4.7}
\eea
where $\cC_{\a(n)}$ is an unconstrained field, and the Lagrange multiplier is 
\bea
G_{\a(n)} =\pa_{(\a_1\a_2} \J_{\a_3 \dots \a_n)}~.
\eea
Varying $S[\cC, G] $ with respect to $\J_{\a(n-2)}$ implies $\cC_{\a(n)} =C_{\a(n)}$,
and then $S[\cC, G] $ reduces to the original action \eqref{4.6}.
On the other hand, we can first integrate out the auxiliary field $\cC_{\a(n)}$ from 
 \eqref{4.7}, which leads to a dual action of the form
\bea
S[ G] =  \int {\rm d}^3x \, 
\vf^6 L \left(\vf^{n-4} G_{\a(n)} \right)~.
\eea
In the $n=2$ case, the duality transformation described corresponds to the standard 
vector-scalar duality in three dimensions.
\\

\noindent
{\bf Acknowledgements:}\\
I am grateful to Mirian Tsulaia for collaboration at the early stage of this project, 
 Jim Gates for important correspondence, Ulf Lindstr\"om for encouragement,  
Joseph Novak for comments on the manuscript, and Paul Townsend
for informing me of the Cottino tensor construction given in \cite{ABdeRST}. 
It is my pleasure to thank the Galileo Galilei Institute for Theoretical Physics 
for the hospitality and the INFN for partial support during the completion of 
the revised version of this work.
This work is supported in part by the Australian Research Council,
project No. DP160103633.

%%%%%%%%%%%%%%%%%%
%%%%%%%%%%%%%%%%%%%

\begin{footnotesize}

\end{footnotesize}

\end{document}